\documentclass{elsarticle}

\usepackage{lineno}

\journal{Physics of Fluids}

\usepackage{graphicx}
\usepackage{amsfonts}
\usepackage{amssymb}
\usepackage{float}
\usepackage{latexsym}
\usepackage{amsmath}

\usepackage{subfigure}
\usepackage[euler]{textgreek}
\usepackage{textcomp}
\usepackage{CJK}

\usepackage{algorithm}
\PassOptionsToPackage{noend}{algpseudocode}
\usepackage[end]{algpseudocode}

\usepackage[table]{xcolor}

\graphicspath{ {./figure/}}
\begin{document}

\begin{frontmatter}

\title{JAX-Shock: A Differentiable, GPU-Accelerated, Shock-Capturing Neural Solver for Compressible Flow Simulation}

\author[add1]{Bo Zhang\corref{cor1}}
\ead{bzhang@niu.edu}
\cortext[cor1]{Corresponding author. 
	Address:  
 	DeKalb, IL 60115, USA
 }
\address[add1]{Department of Mechanical Engineering, Northern Illinois University, DeKalb, IL 60115, USA}


\begin{abstract}
Understanding shock-solid interactions remains a central challenge in compressible fluid dynamics. 
We present JAX-Shock: a fully-differentiable, GPU-accelerated, high-order shock-capturing solver for efficient simulation of the compressible Navier-Stokes equations. 
Built entirely in JAX, the framework leverages automatic differentiation to enable gradient-based optimization, parameter inference, and end-to-end training of deep learning-augmented models. 
The solver integrates fifth-order WENO reconstruction with an HLLC flux to resolve shocks and discontinuities with high fidelity. 
To handle complex geometries, an immersed boundary method is implemented for accurate representation of solid interfaces within the compressible flow field.
In addition, we introduce a neural flux module trained to augment the numerical fluxes with data-driven corrections, significantly improving accuracy and generalization. 
JAX-Shock also supports sequence-to-sequence learning for shock interaction prediction and reverse-mode inference to identify key physical parameters from data. 
Compared with purely data-driven approaches, JAX-Shock enhances generalization while preserving physical consistency. 
The framework establishes a flexible platform for differentiable physics, learning-based modeling, and inverse design in compressible flow regimes dominated by complex shock-solid interactions. 
\end{abstract}


\end{frontmatter}

\section{Introduction}
Simulation of complex shock-solid interactions described by the compressible Euler equations lies at the core of defense, aerospace, and physical sciences, with broad applications ranging from missile aerodynamics~\cite{Tembhurnikar_2025a}, scramjet propulsion~\cite{Kumar_2025a}, and supersonic vehicle design~\cite{Xiao_2018a} to inertial confinement fusion~\cite{Olson_2003a} and astrophysical phenomena such as supernova explosions~\cite{Muller_2020a}. 
Accurately capturing these interactions remains a central challenge in computational fluid dynamics (CFD) due to the coexistence of strong discontinuities, complex flow structures, and intricate coupling between fluid and solid boundaries. 
High-order shock-capturing methods, such as weighted essentially non-oscillatory (WENO) schemes coupled with approximate Riemann solvers~\cite{Johnsen_2006a, Chen_2025a}, have substantially advanced the predictive capability of CFD for shock-dominated flows.
However, the flux formulation and artificial viscosity~\cite{Zhang_2022a}, which are critical for stabilizig numerical shocks, have long stymied progress toward developing robust and differentiable numerical solvers. 
The inherent nonlinearity and discontinuity of these components make gradient-based optimization and sensitivity analysis particularly challenging. 
To address these limitations, this paper introduces a differentiable, GPU-accelerated, shock-capturing neural solver for compressible flow simulation--establishing a flexible platform for differentiable physics, learning-based modeling, inverse design, and high-performance simulation in shock-dominated regimes. 

Classical, solver-free deep learning approaches learn mappings between finite-dimensional Euclidean spaces through neural networks~\cite{Zhang_2023b, Zhang_2023a} and extend this capability to infinite-dimensional Banach spaces of functions via neural operators~\cite{Zhang_2025a}. 
In contrast, differentiable programming provides a unifying framework that bridges scientific computing and machine learning (ML)~\cite{Innes_2019a}, enabling the seamless integration of conventional numerical solvers with end-to-end trainable ML architectures. 
This integration allows for gradient-based optimization of physical parameters, data assimilation, and discovery of governing equations directly from simulation or experimental data. 
Notably, physics-informed neural networks (PINNs)~\cite{Raissi_2019a} have emerged as a key instantiation of differentiable programming, embedding physical laws as soft constraints into the loss function to enable solver-free, physics-constrained learning. 
Among the state-of-the-art tools for automatic differentiation (AD) in Python, TensorFlow~\cite{Abadi_2016a}, PyTorch~\cite{Paszke_2019a}, and JAX~\cite{jax2018github} have emerged as the dominant frameworks. 
While TensorFlow and PyTorch are widely adopted in the ML community, JAX has gained particular traction in scientific computing due to its composable function transformations (\textit{e.g.}, grad, vmap, and pmap) and just-in-time (JIT) compilation through accelerated linear algebra (XLA), which together enable high-performance, GPU-accelerated, and fully differentiable numerical simulations. 

A flurry of recent studies has focused on developing differentiable hybrid neural solvers that integrate ML models with traditional numerical simulation frameworks, demonstrating their potential for scalable, physics-based differentiable computing.  
This paradigm has been exemplified through the gradient-based end-to-end optimization across a wide range of domains, including fluid dynamics~\cite{Sirignano_2020a, Kochkov_2021a, List_2022a, Bezgin_2023a, Zhang_2023c}, as well as other fields such as the finite element method~\cite{Xue_2023a}, molecular dynamics~\cite{Schoenholz_2021a}, nanoscale heat transfer~\cite{Zhang_2024a}, and density functional theory~\cite{Li_2021a}. 
In the realm of fluid dynamics, Sirignano \textit{et al.}~\cite{Sirignano_2020a} leveraged a neural network to learn unknown physics from data and augment the governing partial differential equation. 
In contrast to purely black-box ML approaches, Kochkov \textit{et al.}~\cite{Kochkov_2021a} modeled two-dimenional turbulent flows using an end-to-end differentiable framework, achieving substantial computational speedups and strong generalization to unseen flow regimes. 
List \textit{et al.}~\cite{List_2022a} further developed a differentiable numerical solver that enables the propagation of optimization gradients through multiple solver steps, allowing turbulence models to be trained to improve under-resolved, low-resolution solutions to the incompressible Navier-Stokes equations. 
Bezgin \textit{et al.}~\cite{Bezgin_2023a} introduced JAX-Fluids, a comprehensive, fully-differentiable CFD solver for compressible two-phase flows, enabling end-to-end optimization and seamless hybridization of ML with CFD. 

Despite these advances, a differentiable neural solver capable of simulating shock-solid interactions via an immersed boundary method (IBM) remains unexplored. 
To the best of our knowledge, this work presents the first differentiable hybrid neural solver that incorporates IBM to simulate compressible flows with shock-solid interactions, marking a significant step toward differentiable physics-based modeling of complex fluid-structure systems. 
We further introduce a neural flux module that augments the numerical fluxes with data-driven corrections, significantly enhancing accuracy and generalization. 

\section{Approach} 
\label{sec:approach}

\begin{figure}
\begin{center}
\includegraphics [width=.95\columnwidth]{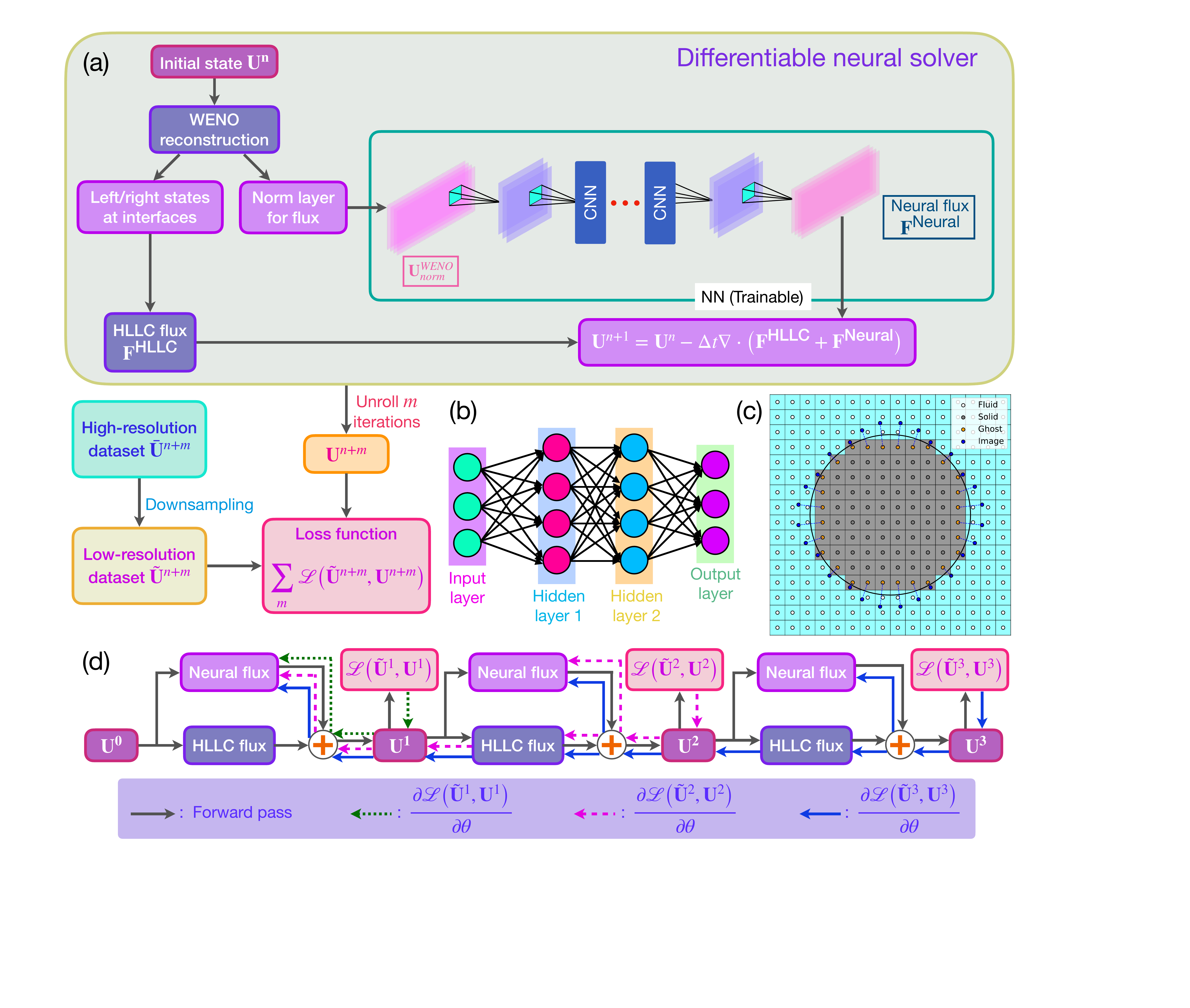}
\end{center}
\caption{Overview of our differentiable framework. (a) The solver workflow of the JAX-Shock framework for two-dimensional simulations, integrated with a series of CNNs for neural flux computation. The solver unrolls for $m$ time steps. The initial state is passed through the framework to obtain the neural flux after WENO reconstruction and a normalization layer, and the final flux is formed by combining the HLLC and neural fluxes. High-resolution reference data are downsampled to generate low-resolution targets, which are compared with the solver outputs to compute the loss. (b) A multilayer perceptron (MLP) network used for the one-dimensional Sod shock-tube simulation to predict the neural flux. (c) Schematic of the node classification for fluid, solid, ghost, and image cells in the immersed boundary method. (d) Visualization of gradient backpropagation in a 3-step setup, where the loss gradients from the final step are propagated through all preceding steps and corresponding network outputs.}
\label{fig:solver}
\end{figure}  

In this paper, a differentiable hybrid neural solver is developed to study shock-solid interactions in compressible fluids. 
The Euler equations govern such compressible inviscid flows, written here in two dimensions for generality as 
\begin{align}
	\frac{\partial \mathbf{U}}{\partial t} + \nabla \cdot \mathbf{F}(\mathbf{U}) & = 0 \, ,
 	 \label{eq:euler}
\end{align}
where the vector of conservative variables and the flux tensor are defined as
\[
\mathbf{U} = 
\begin{bmatrix}
\rho \\
\rho u \\
\rho v \\
E
\end{bmatrix}, \quad
\mathbf{F}(\mathbf{U}) =
\begin{bmatrix}
\rho u & \rho v \\
\rho u^2 + p & \rho u v \\
\rho u v & \rho v^2 + p \\
u(E+p) & v(E+p)
\end{bmatrix},
\]
with \( \rho \), \( u \), \( v \), \( E \), and \( p \) denoting the density, velocity components, total energy, and pressure, respectively. 
The system is closed by the equation of state,
\begin{align}
p &= (\gamma - 1)\left(E - \frac{1}{2}\rho(u^2 + v^2)\right),
\end{align}
where \( \gamma \) is the specific heat ratio.

High-order accurate WENO schemes with convex reconstruction of candidate stencils are employed in this work, in conjunction with the positivity-preserving HLLC (Harten-Lax-van Leer-Contact) approximate Riemann solver~\cite{Johnsen_2006a}, to achieve robust shock-capturing.
This combination enables accurate resolution of discontinuities with correct wave speeds in single-fluid Riemann problems. 
Primitive variables are reconstructed within a finite volume framework that naturally accommodates the staggered discretization to ensure smooth and stable advection without introducing spurious oscillations. 
The HLLC flux along the $x$-direction is expressed as
\begin{align}
\mathbf{F}^{\text{HLLC}}_x &= \frac{1+\text{sign}(s^*)}{2}[\mathbf{F}^{\text{L}}+s^{-}(\mathbf{U}^{\text{*L}}-\mathbf{U}^{\text{L}})] + \frac{1-\text{sign}(s^*)}{2}[\mathbf{F}^{\text{R}}+s^{+}(\mathbf{U}^{\text{*R}}-\mathbf{U}^{\text{R}})],
\end{align}
where the left and right states ($k =$ L, R) are obtained via WENO reconstruction, and the intermediate (star) states are defined as
\begin{align}
\mathbf{U}^{\text{*k}} &= \frac{s^k-u^k}{s^k-s^*}
\begin{bmatrix}
\rho^{k} \\
\rho^{k} s^* \\
\rho^{k} v^{k} \\
E^k \frac{s^k-s^*}{s^k-u^k}+(s^*-u^{k})\bigg(\rho^{k} s^* - \rho^{k} u^{k} \frac{s^k-s^*}{s^k-u^k}\bigg)
\end{bmatrix}. 
\end{align}
The wave speeds in the HLLC solver are defined as 
\begin{align}
s^{+}=\max(0,s^{\text{R}}), \quad s^{-}=\min(0,s^{\text{L}})
\end{align}
where the left- and right-going waves are estimated by
\begin{align}
s^{\text{L}} = \min(u^{\text{L}}-a^{\text{L}},u^{\text{R}}-a^{\text{R}}), \quad s^{\text{R}} = \max(u^{\text{L}}+a^{\text{L}},u^{\text{R}}+a^{\text{R}})
\end{align}
where $a^{k}=\sqrt{\gamma p^{k}/\rho^{k}}$ is the speed of sound. 
The contact wave speed is computed as
\begin{align}
s^{*} &= \frac{p^{\text{R}}-p^{\text{L}} + \rho^{\text{L}}u^{\text{L}}(s^{\text{L}}-u^{\text{L}}) - \rho^{\text{R}}u^{\text{R}}(s^{\text{R}}-u^{\text{R}})}{\rho^{\text{L}}(s^{\text{L}}-u^{\text{L}})-\rho^{\text{R}}(s^{\text{R}}-u^{\text{R}})}. 
\end{align}
Analogously, the HLLC flux along the $y$-direction, $\mathbf{F}^{\text{HLLC}}_y$, is obtained by interchanging the $x$ and $y$ velocity components. 
The total HLLC flux tensor is then assembled as 
\begin{align}
\mathbf{F}^{\text{HLLC}} &= 
\begin{bmatrix}
\mathbf{F}^{\text{HLLC}}_x & \mathbf{F}^{\text{HLLC}}_y
\end{bmatrix}.
\end{align}
In cell-face notation, the numerical fluxes are expressed as
\begin{align}
\mathbf{F}^{\text{HLLC}}_x = \text{HLLC}(\mathbf{U}^{\text{L}}_{i+1/2,j}, \mathbf{U}^{\text{R}}_{i+1/2,j}),  \quad \mathbf{F}^{\text{HLLC}}_y = \text{HLLC}(\mathbf{U}^{\text{L}}_{i,j+1/2}, \mathbf{U}^{\text{R}}_{i,j+1/2}),
\end{align} 
which represent the HLLC fluxes evaluated at the faces $x_{i+1/2}$ and $y_{j+1/2}$ of cell $(i,j)$.
To further enhance numerical stability, a Laplacian-type artificial viscosity term, $\nu \nabla^2 \mathbf{U}$, is added to the conservative update, 
\begin{align}
\mathbf{U}^{n+1} &= \mathbf{U}^{n} - \Delta t \, \nabla \cdot \mathbf{F}^{\text{HLLC}}(\mathbf{U}^{\text{WENO}}) + \Delta t \, \nu \, \nabla^2 \mathbf{U}^{n}
\end{align}
where $\mathbf{U}^{n}$ denotes the conservative variables at time step $n$, $\mathbf{U}^{n+1}$ the updated solution, $\mathbf{U}^{\text{WENO}}$ the WENO-reconstructed conservative variables from $\mathbf{U}^{n}$, $\Delta t$ the time step size, and $\nu$ a small numerical diffusion coefficient. 
This artificial viscosity term effectively suppresses unphysical oscillations near discontinuities while preserving accuracy in smooth regions. 

The JAX-Shock framework operates in two modes.  
The first mode serves as a fully differentiable solver that enables gradient-based parameter optimization without neural network training. 
This is achieved by implementing the entire solver in JAX, which leverages automatic differentiation and just-in-time compilation for efficient, GPU-accelerated computation.
The second mode integrates a neural flux module to enhance low-resolution simulations, enabling accurate reconstruction of high-resolution flow fields through end-to-end learning. 
An overview of the solver workflow incorporating the neural flux module is illustrated in figure~\ref{fig:solver}(a). 

For two-dimensional simulations, the neural flux model is parameterized by a fully convolutional neural network (CNN) consisting of three convolutional layers. 
The first two layers employ rectified linear unit (ReLU) activations, while the final layer uses a hyperbolic tangent (tanh) activation. 
The network contains a total of 6,388 trainable parameters and preserves spatial resolution via ``SAME" padding. 
The CNN is specifically designed for pixel-wise prediction of the neural flux field, ensuring local spatial consistency across the computational domain. 
As shown in figure~\ref{fig:solver}(a), the neural flux module takes as input the normalized WENO-reconstructed states, denoted by $\mathbf{U}^{\text{WENO}}_{\text{norm}}$, and interacts with the numerical flux computation within each update step.
This formulation allows the learned flux correction to refine the baseline HLLC flux, thereby improving the overall solution accuracy. 
The conservative update equation is expressed as 
\begin{align}
\mathbf{U}^{n+1} &= \mathbf{U}^{n} - \Delta t \, \nabla \cdot (\mathbf{F}^{\text{HLLC}}(\mathbf{U}^{\text{WENO}}) + \mathbf{F}^{\text{Neural}}(\mathbf{U}^{\text{WENO}}_{\text{norm}} \mid \theta)) + \Delta t \, \nu \, \nabla^2 \mathbf{U}^{n}
\end{align}
where $\mathbf{F}^{\text{Neural}}(\mathbf{U}^{\text{WENO}}_{\text{norm}}) : \mathbb{R}^{\tilde{N}_x \times \tilde{N}_y \times 4} \;\longrightarrow\; \mathbb{R}^{\tilde{N}_x \times \tilde{N}_y \times 4}$ is defined over the flow field of size $\tilde{N}_x \times \tilde{N}_y$ and denotes the neural flux obtained from the neural flux operator $\mathcal{F}^{\text{Neural}}$, parameterized by the learnable network weights $\theta$: 
\begin{align}
\mathcal{F}^{\text{Neural}} : \mathbf{U}^{\text{WENO}}_{\text{norm}} \mapsto \mathbf{F}^{\text{Neural}}(\mathbf{U}^{\text{WENO}}_{\text{norm}} \mid \theta).
\end{align}
The total neural flux is composed of its directional components,
\begin{align}
\mathbf{F}^{\text{Neural}}(\mathbf{U}^{\text{WENO}}_{\text{norm}} \mid \theta) = \begin{bmatrix}
\mathbf{F}^{\text{Neural}}_x(\mathbf{U}^{\text{WENO}_x}_{\text{norm}} \mid \theta) & \mathbf{F}^{\text{Neural}}_y(\mathbf{U}^{\text{WENO}_y}_{\text{norm}} \mid \theta)
\end{bmatrix}
\end{align}
where $\mathbf{U}^{\text{WENO}_{x}}_{\text{norm}}$ and $\mathbf{U}^{\text{WENO}_{y}}_{\text{norm}}$ represent the normalized conservative variables reconstructed along the $x$ and $y$ directions, respectively.
Both components are generated by CNNs sharing the same set of learnable parameters $\theta$. 
For the one-dimensional Sod shock-tube simulation, a multilayer perceptron (MLP) network is used to predict the neural flux, see figure~\ref{fig:solver}(b). 
The ghost cell immersed boundary method (IBM)~\cite{Mittal_2005a} is adopted to handle solid boundaries within a Cartesian grid framework, as shown in figure~\ref{fig:solver}(c). 
A slip-wall boundary condition is implemented, enforcing zero normal penetration while preserving the tangential component of velocity. 
Computational cells are classified as fluid, solid, ghost, or image cells. 

The proposed neural flux operator is designed to recover high-resolution flow features from simulations performed on a coarse grid. 
High-resolution reference data are downsampled through a block-averaging procedure $d(\mathbf{U}): \mathbb{R}^{N_x \times N_y \times 4} \rightarrow \mathbb{R}^{\tilde{N}_x \times \tilde{N}_y \times 4}$ to construct the corresponding low-resolution training targets $\{\tilde{\mathbf{U}}^{n+m}\}$ after unrolling $m$ steps of the JAX-Shock solver. 
During training over these $m$ unrolled steps, the optimization objective is formulated as the mean squared error between the predicted and target flow states: 
\begin{align}
\mathcal{L}(\theta) = \frac{1}{m}\sum_{s=1}^{m} \mathcal{L}_s(\theta)
= \frac{1}{m}\sum_{s=1}^{m} \big\| \mathbf{U}^{\,n+s}(\theta) - \tilde{\mathbf{U}}^{\,n+s} \big\|_2^2.
\end{align}
where $\mathcal{L}_s(\theta)$ is the loss at an intermediate solver step, $\mathbf{U}^{n+s}(\theta)$ denotes the solver prediction parameterized by the neural flux network with weights $\theta$, and $\tilde{\mathbf{U}}^{n+s} = d(\mathbf{U}_{\text{HR}}^{n+s})$ represents the downsampled low-resolution targets derived from the high-resolution ground-truth state $\mathbf{U}_{\text{HR}}^{n+s}$. 
During training, gradients of all intermediate losses $\mathcal{L}_s(\theta)$ with respect to neural flux network parameters $\theta$ are computed via backpropagation through the full sequence of unrolled solver steps, as illustrated in figure~\ref{fig:solver}(d). 
Let the solver update be written as
\begin{align}
\mathbf{U}^{n+k}(\theta) &= \mathcal{S}\big(\mathbf{U}^{n+k-1}(\theta),\, \theta\big), \quad k = 1, \ldots, m, 
\end{align}
where $\mathcal{S}$ denotes a single solver step incorporating the neural flux. 
Then, the gradient of an intermediate loss with respect to $\theta$ can be formally expressed using the chain rule:  
\begin{align}
\frac{\partial \mathcal{L}_s}{\partial \theta} 
&=
\sum_{k=1}^{s} 
\frac{\partial \mathcal{L}_s}{\partial \mathbf{U}^{n+s}} 
\left(
\prod_{j=k+1}^{s} 
\frac{\partial \mathbf{U}^{n+j}}{\partial \mathbf{U}^{n+j-1}}
\right)
\frac{\partial \mathbf{U}^{n+k}}{\partial \mathbf{F}_{k-1}^{\text{Neural}}}
\frac{\partial \mathbf{F}_{k-1}^{\text{Neural}}}{\partial \theta}.
\end{align}
where $\mathbf{F}_{k}^{\text{Neural}}$ represents the neural flux at step $k$~\cite{List_2022a}. 

\section{Results}
\label{sec:results}

\begin{figure}
\begin{center}
\includegraphics [width=.8\columnwidth]{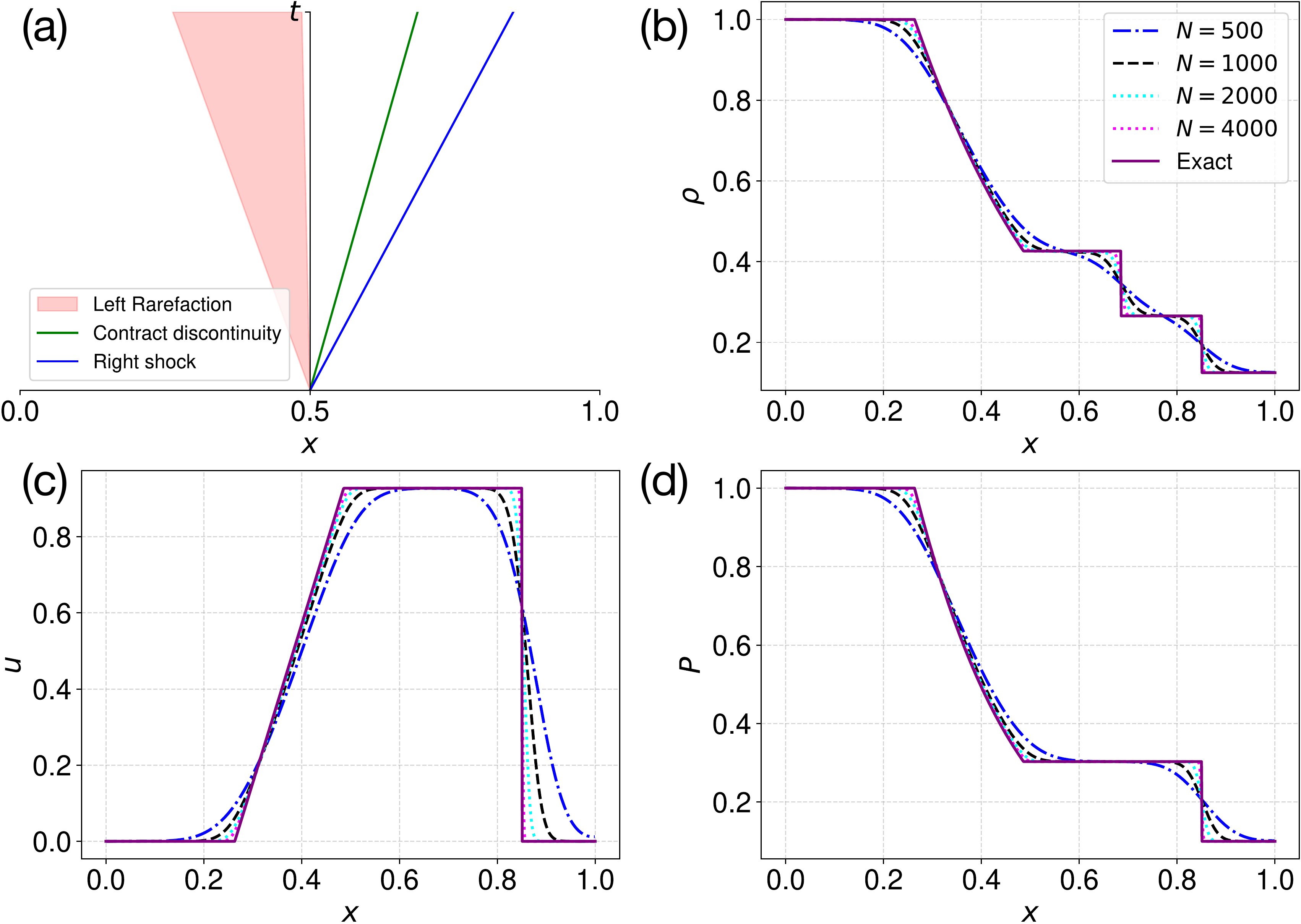}
\end{center}
\caption{Sod problem at $t=0.2$: (a) wave structure, (b) density, (c) velocity, and (d) pressure computed with the present neural solver. The solid line represents the exact solution, while the other lines depict results for different mesh resolutions.}
\label{fig:shock_sod}
\end{figure}

\begin{figure}
\begin{center}
\includegraphics [width=1\columnwidth]{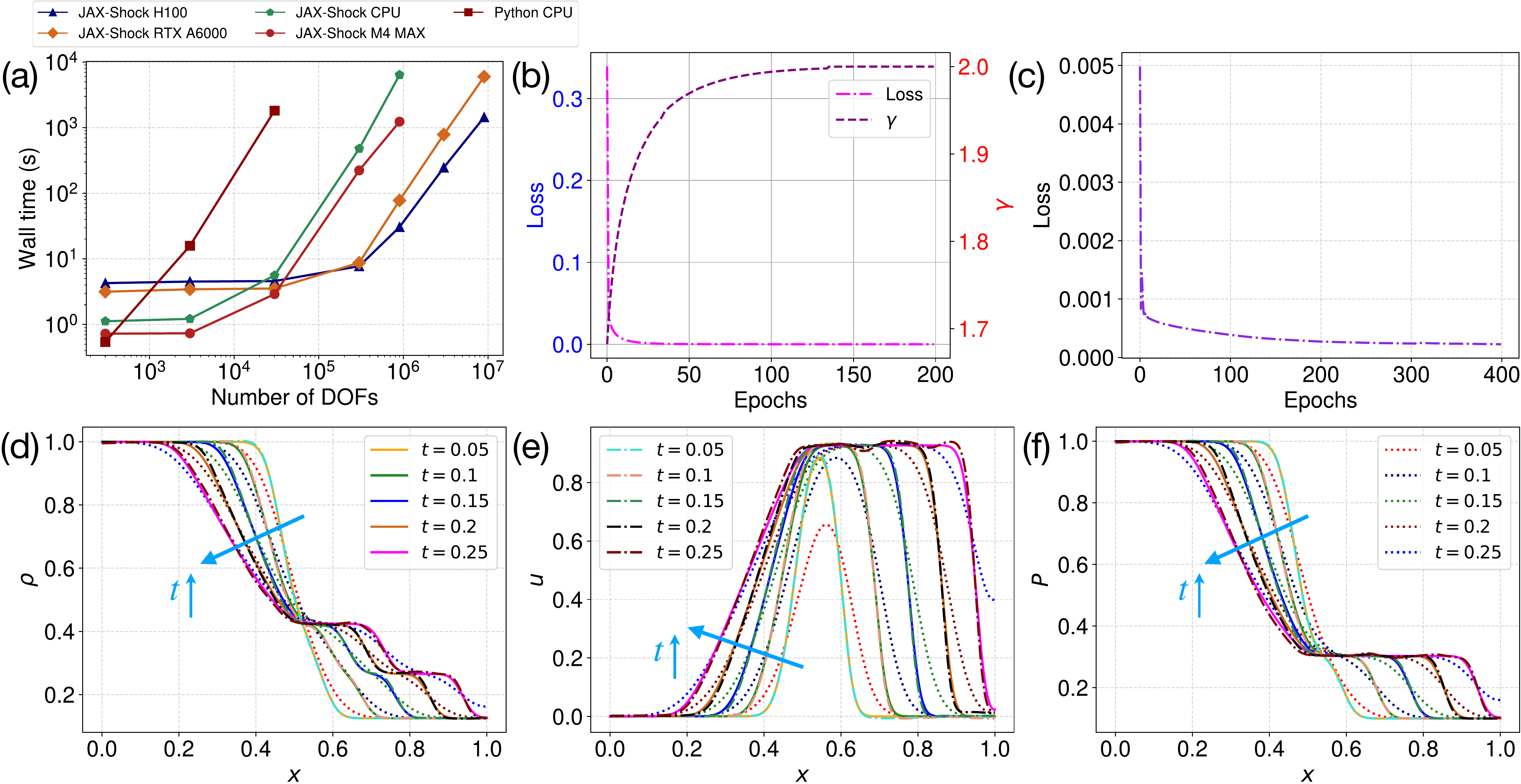}
\end{center}
\caption{Sod shock tube problem. (a) Computing performance and scalability of JAX-Shock, showing wall time as a function of the number of DOFs meansured on the one-dimensional Sod problem. (b) Reverse learning (optimization) of specific heat ratio, $\gamma$, and the corresponding training loss history. (c) Training loss history for learning the neural flux operator. Temporal evolution of (d) density, (e) velocity, and (f) pressure profiles: neural-flux enhanced low-resolution solution (($N=500$, dash-dotted), high-resolution reference ($N=1000$, solid), and baseline low-resolution solution without neural flux ($N=500$, dotted).}
\label{fig:shock_sod_nn}
\end{figure}

The Sod shock-tube problem~\cite{Sod_1978a} serves as a canonical benchmark for assessing the performance of compressible flow solvers, offering an idealized yet rigorous test of convective fluxes formulation, shock-capturing accuracy and wave propagation fidelity. 
The initial conditions for the left and right states are given by
\begin{align}
(\rho, u, P, \gamma)^{\text{T}}_{\text{L}} = (1, 0, 1, 1.4)^{\text{T}}_{\text{L}}, \quad (\rho, u, P, \gamma)^{\text{T}}_{\text{R}} = (0.125, 0, 0.1, 1.4)^{\text{T}}_{\text{R}}.  
\end{align} 
As shown in figure~\ref{fig:shock_sod}(a), three distinct waves emanate from the initial discontinuity: a left-propagating rarefaction wave, a right-propagating contact discontinuity, and a right-propagating shock. 
As the spatial resolution $N$ is increased, the numerical solution converges toward the exact solution. 
Moreover, the computed solution aligns very well with the analytical solution, as illustrated in figures~\ref{fig:shock_sod}(b)-(d).

The performance and scalability of JAX-Shock are benchmarked using the Sod problem under various mesh resolutions across multiple computational architectures, including an Intel(R) Core(TM) Ultra 7 165U CPU @ 2.05GHz under Windows operating system, an Apple M4 Max GPU (40-core with 128GB unified memory) on macOS Sequoia 15.5, and NVIDIA GPUs (RTX A6000 with 48 GB and H100 Hopper NVL with 94 GB graphics memory) on Ubuntu 22.04.5 LTS.  
Figure~\ref{fig:shock_sod_nn}(a) presents the wall-clock time as a function of the number of degrees of freedom (DOFs). 
JAX-Shock demonstrates a significant performance advantage on GPUs compared with CPU execution.
The largest problem, consisting of $9 \times 10^6$ DOFs, requires 5992 s on an RTX A6000 and 1445 s on an H100, yielding a 4.1$\times$ speedup for the H100 over the A6000.
In contrast, the largest feasible cases on CPU and Apple M4 MAX configurations contain $9 \times 10^5$ DOFs and take 6375 s and 1234 s, respectively. 
For comparison, a naive Python implementation using explicit for-loops (without JAX) takes 1822 s on CPU for only $3 \times 10^4$ DOFs. 
Relative to this baseline, the H100 and A6000 achieve speedups of 396.5$\times$ and 515.1$\times$, respectively. 
When compared with JAX implementations on CPU and M4 MAX, the H100 achieves 208$\times$ and 40.2$\times$ speedups, respectively, for simulations with $9 \times 10^5$ DOFs. 
Figure~\ref{fig:shock_sod_nn}(b) illustrates the reverse learning (optimization) process used to infer the specific heat ratio, $\gamma$, from the Sod shock-tube data using a constant learning rate of $0.1$.
In this experiment, $\gamma$ is treated as a learnable parameter within the differentiable solver and is iteratively updated through gradient-based optimization to minimize the $L_2$ loss between the predicted and reference solutions.
The figure presents both the evolution of the estimated $\gamma$ toward its true value and the corresponding training loss history, demonstrating rapid convergence and confirming the capability of JAX-Shock to recover underlying physical parameters directly from flow field data. 
The neural flux module in JAX-Shock is trained using low-resolution simulations with $N=500$ grid points. 
The reverse learning and neural flux models are all trained using the Adam optimizer. 
As shown in figure~\ref{fig:shock_sod_nn}(c), the training loss converges rapidly when using a constant learning rate schedule with a learning rate of $10^{-3}$. 
After incorporating the neural flux module, the temporal evolution of density, velocity and pressure profiles obtained from the forward pass of the low-resolution simulation agrees very well with the high-resolution reference results ($N=1000$) and substantially outperforms the baseline low-resolution simulation ($N=500$), as illustrated in figures~\ref{fig:shock_sod_nn}(d)-(f). 
The generalizability to unseen future times is also evaluated. 
Strikingly, the neural flux model is trained only up to the final time $t = 0.2$, yet JAX-Shock maintains accurate predictions when extrapolated to $t = 0.25$, as shown in figures~\ref{fig:shock_sod_nn}(d)-(f).

\begin{figure}
\begin{center}
\includegraphics [width=1\columnwidth]{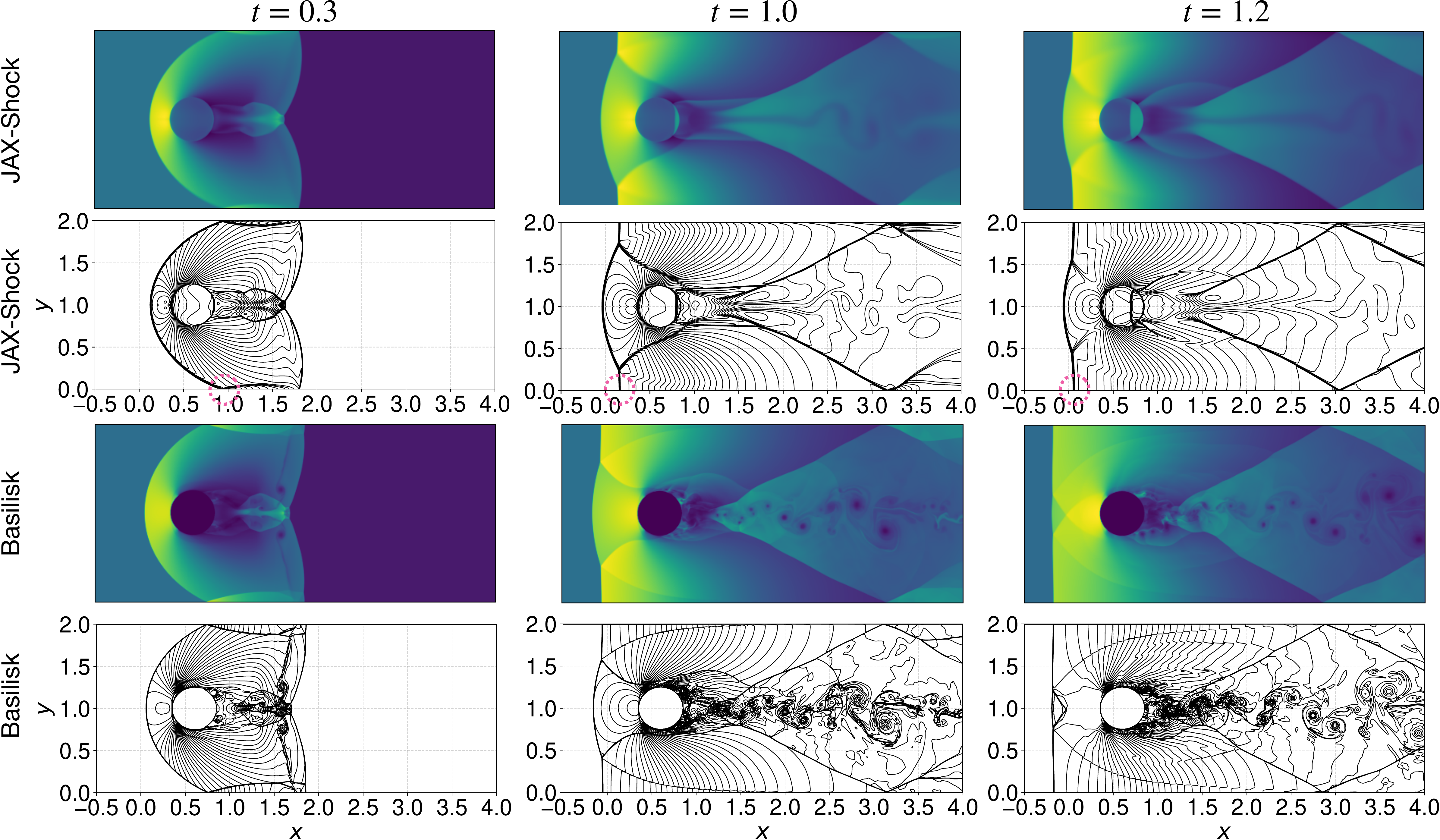}
\end{center}
\caption{Interaction of a shock with a circular cylinder: comparison of density fields computed using JAX-Shock and Basilisk at three time instances, $t = 0.3$, $1.0$, and $1.2$. Each pair of panels shows color contours (top) and black line contours (bottom) with 32 contour levels ranging from 0.8 to 18.}
\label{fig:shock_cylinder_inlet}
\end{figure}

\begin{figure}
\begin{center}
\includegraphics [width=1\columnwidth]{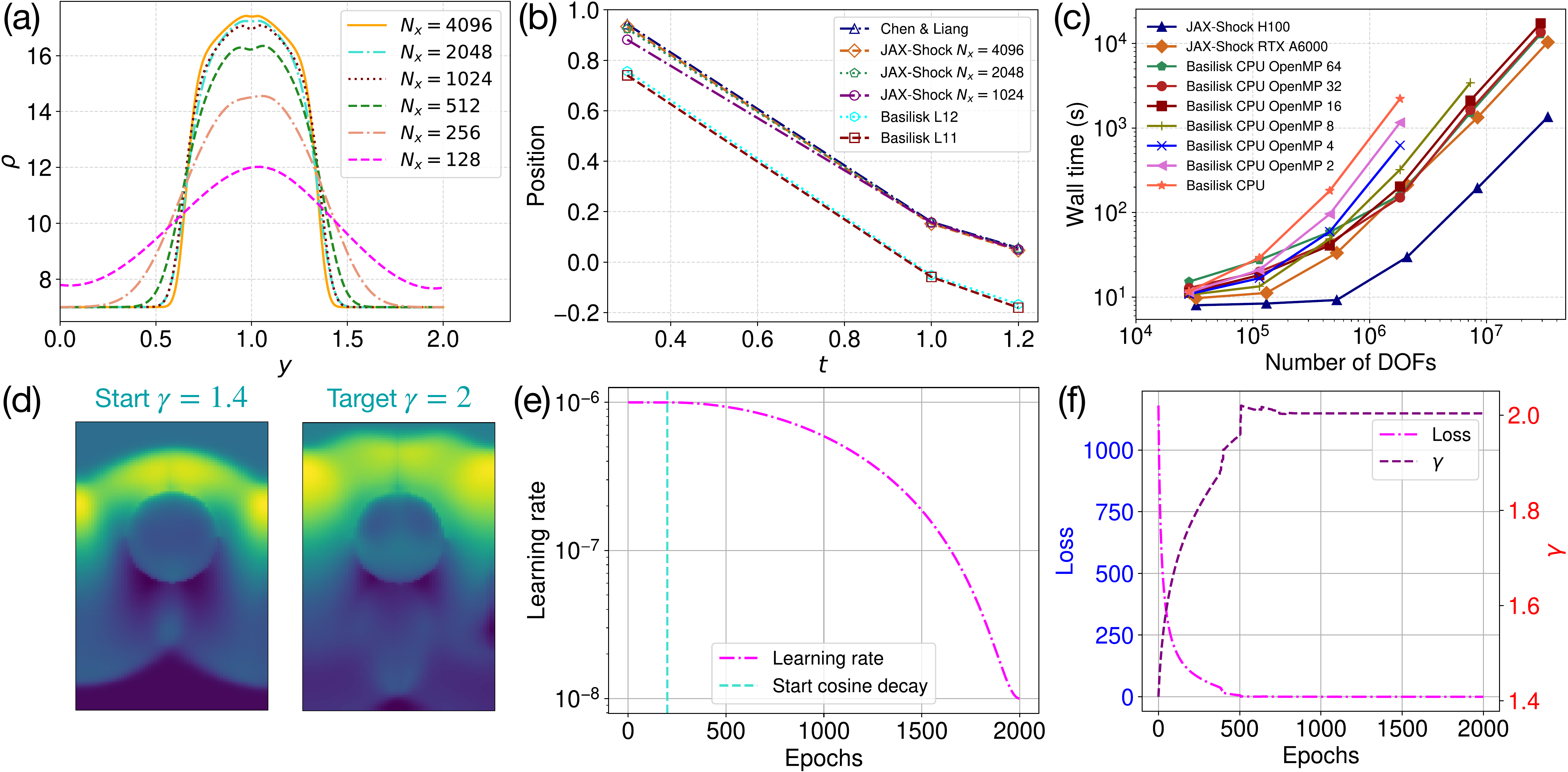}
\end{center}
\caption{Interaction of a moving normal shock with a circular cylinder. (a) Convergence study based on the line profiles of density along the transverse direction at $x=0.2$ and $t=0.3$. (b) Validation of JAX-Shock by comparing the temporal evolution of the intersection point between the incident shock and the bottom wall (see dashed circles in figure~\ref{fig:shock_cylinder_inlet}) across multiple grid resolutions, benchmarked against Basilisk and reference data~\cite{Chen_2025a}. (c) Computational performance and scalability of JAX-Shock, showing wall time as a function of the number of DOFs meansured for the shock-cylinder interaction. (d) Restricted reverse-learning domain for the density fields, illustrating the initial guess ($\gamma=1.4$) and target value ($\gamma=2$). (e) Learning rate annealing schedule used in the reverse-learning process. (f) Reverse learning (optimization) of the specific heat ratio $\gamma$ and the corresponding training loss history.}
\label{fig:shock_cylinder_inlet_nn}
\end{figure}

In the next experiment, the interaction of a moving normal shock with a circular cylinder of radius 0.25, centered at $(0.6,1)$, is simulated within a two-dimensional computational domain spanning $(x,y) \in [-0.5,4] \times [0,2]$~\cite{Chen_2025a}. 
The flow field is initialized with high-density and high-pressure upstream values, creating a sharp normal shock with a speed of 5, located to the left of the cylinder, while the downstream region is set to low-density, low-pressure conditions, specifically: 
\begin{align}
 \{\rho, u, v, p, \gamma\} &=
 \begin{cases}
\{7, 4, 0, 29, 1.4\} & -0.5 \le x < 0.3, \\
  \{1.4, 0, 0, 1, 1.4\} & 0.3 \le x \le 4.
 \end{cases}
 \label{eq:Sod_ST}
\end{align}
The left boundary is prescribed as an inflow condition consistent with the left state of the initial flow, while the right boundary is prescribed as an outflow condition.
The top and bottom boundaries are treated as reflective walls. 
Figure~\ref{fig:shock_cylinder_inlet} presents the computed density fields obtained using JAX-Shock and Basilisk~\cite{Popinet_2009a} at three time instants, $t = 0.3$, $1.0$, and $1.2$. 
The Basilisk simulation employs an $L12$ mesh resolution without mesh adaption, comprising $\sim$$7.3 \times 10^6$ cells, while the JAX-Shock computation uses a uniform grid of $\sim$$8.4 \times 10^6$ cells. 
Both solvers capture the complex interaction between the incident shock and cylinder, including the formation of the incident, reflected, and transmitted shocks, as well as the subsequent wake dynamics. 
At $t = 1.0$ and $t=1.2$, two well-defined shock wave triple points arise symmetrically along the $x$-axis and are clearly visible in both solutions.  
However, the vortex shedding behind the cylinder appears less pronounced in JAX-Shock compared with Basilisk, despite the former employing a high-order WENO5 reconstruction and HLLC flux scheme, whereas Basilisk uses the second-order Bell-Collela-Glaz unwind scheme with a minmod slope limiter. 
Although WENO5 is formally fifth-order accurate in smooth regions, the additional artificial dissipation introduced in JAX-Shock can smear vorticity layers, thereby attenuating fine-scale wake structures.

As shown in figure~\ref{fig:shock_cylinder_inlet_nn}(a), the density profile gradually converges as the mesh is refined. 
Figure~\ref{fig:shock_cylinder_inlet_nn}(b) demonstrates that the temporal evolution of the intersection point between the incident shock and the bottom wall predicted by JAX-Shock aligns well with the reference data across all refined meshes, whereas the Basilisk results exhibit noticeable deviations. 
The computational performance and scalability of JAX-Shock on GPUs are assessed and compared with Basilisk running on CPUs with OpenMP parallelization. 
JAX-Shock is executed on a 2.0 GHz Intel(R) Xeon(R) Gold 6438Y+ CPU (32 cores) and NVIDIA GPUs (RTX A6000, 48 GB; H100 Hopper NVL, 94 GB) on Ubuntu 22.04.5 LTS.
Basilisk with OpenMP runs on the same CPU platform. 
Figure~\ref{fig:shock_cylinder_inlet_nn}(c) shows the wall-clock time as a function of the number of DOFs. 
For the largest problem with $\sim$$3.4 \times 10^7$ DOFs, the RTX A6000 requires 10354 s, while the H100 completes the same case in 1350 s, corresponding to a 7.6$\times$ speedup.
In contrast, the OpenMP acceleration of Basilisk becomes marginal at large DOF counts: 
increasing the number of CPU cores from 16 to 64 yields only a 1.3$\times$ speedup for a simulation with $2.9 \times 10^7$ DOFs.
For comparable problem sizes, the H100 and A6000 achieve speedups of 9.6$\times$ and 1.2$\times$, respectively, over Basilisk OpenMP with 64 cores. 
Relative to the Basilisk CPU case (with $1.8 \times 10^6$ DOFs), the H100 and A6000 deliver 74.4$\times$ and 10.5$\times$ speedups for simulations with $2.1 \times 10^6$ DOFs--despite operating on larger problems. 
Figure~\ref{fig:shock_cylinder_inlet_nn}(d) presents the restricted training domain, defined by $(x,y) \in [0,1.5] \times [0,1]$ and comprising 8192 cells.  This subdomain is used for reverse learning of the specific heat ratio, $\gamma$, starting from an initial guess and iteratively converging toward the target value. 
To accelerate optimization and improve stability, a learning-rate annealing strategy based on a warm cosine decay schedule is employed, see figure~\ref{fig:shock_cylinder_inlet_nn}(e). 
After an initial 200 epochs at a constant learning rate of $10^{-6}$, the cosine decay is activated, reducing the learning rate from $10^{-6}$ to $10^{-8}$. 
As shown in figure~\ref{fig:shock_cylinder_inlet_nn}(f), the training loss decreases rapidly under this annealing schedule, demonstrating stable and efficient convergence of the reverse-learning procedure. 

\begin{figure}
\begin{center}
\includegraphics [width=1\columnwidth]{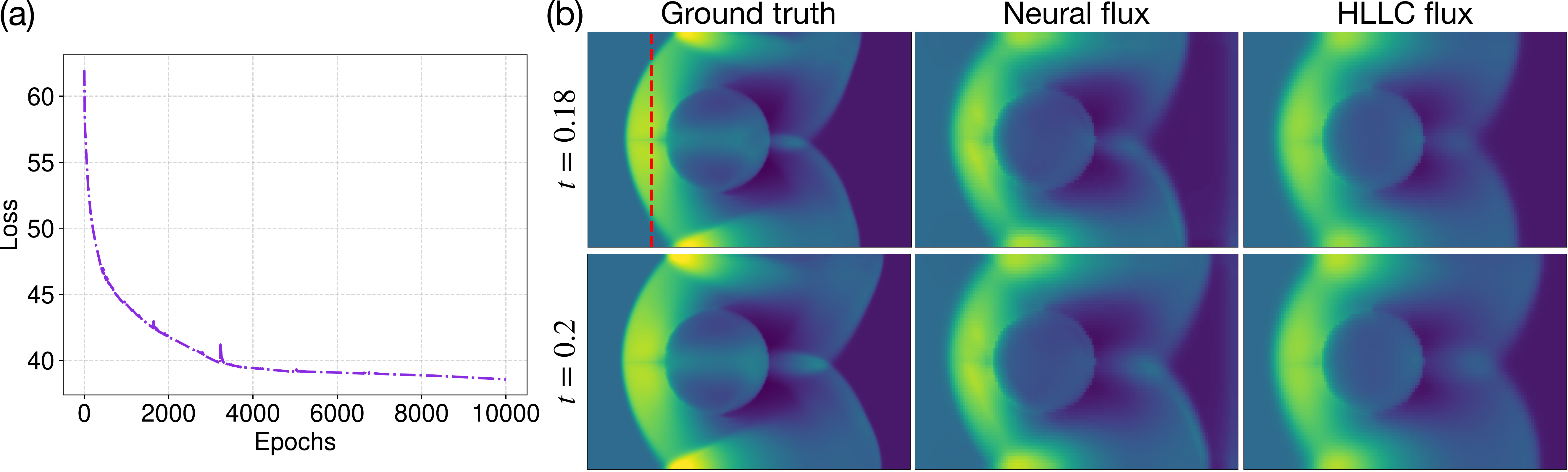}
\end{center}
\caption{Learned neural flux operator for the interaction of a moving normal shock with a circular cylinder. (a) Training loss history for learning the neural flux operator. (b) Evolution of the predicted density fields for reference ($256 \times 128$), learned ($128 \times 64$), and baseline ($128 \times 64$) solvers.}
\label{fig:shock_cylinder_inlet_nf}
\end{figure}

\begin{figure}
\begin{center}
\includegraphics [width=1\columnwidth]{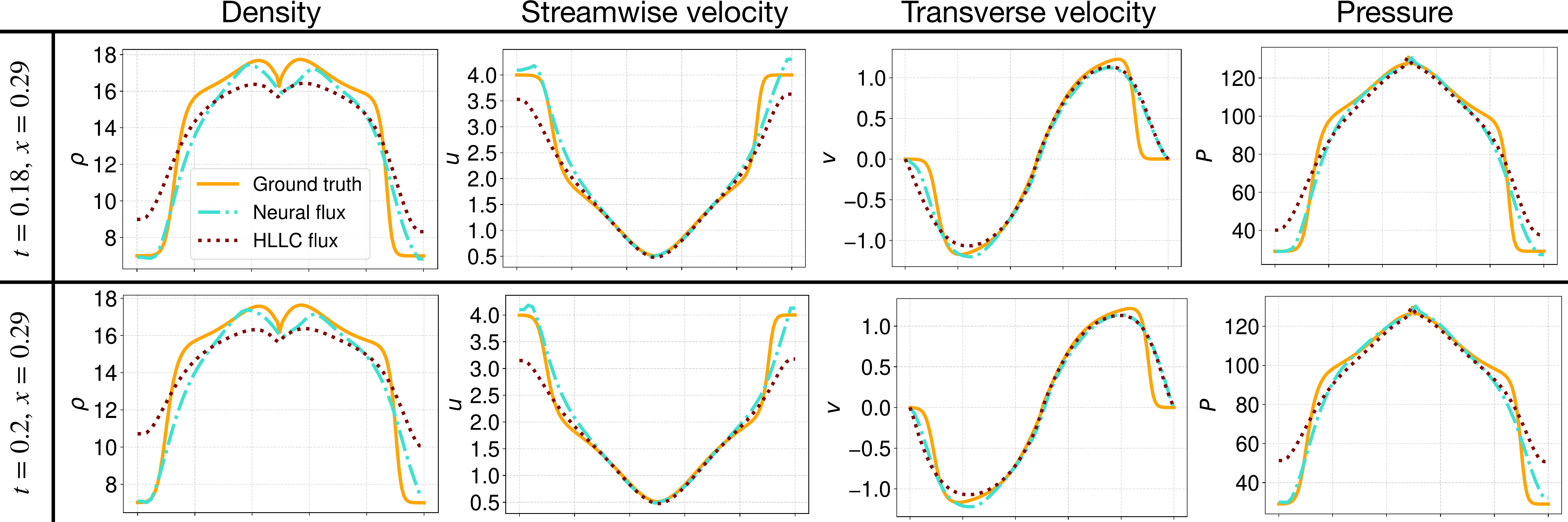}
\end{center}
\caption{Comparison of the learned temporal evolution of the predicted density, velocity and pressure profiles along the $y$-direction against the reference solution and baseline results.}
\label{fig:shock_cylinder_inlet_nf_line}
\end{figure}

The neural flux model is trained using a piecewise constant learning rate schedule, with an initial learning rate of $10^{-4}$ for the first 3500 iterations, followed by a reduced rate of $10^{-5}$ for the remainder of the training.
The resulting training loss history is shown in figure~\ref{fig:shock_cylinder_inlet_nf}(a), where the loss exhibits a gradual decay and convergence after many epochs. 
Figure~\ref{fig:shock_cylinder_inlet_nf}(b) compares the temporal evolution of the density fields predicted by the learned neural flux model against the ground truth and the baseline solver. 
The sharp shock is well captured by the neural flux model at the same resolution ($128 \times 64$) as the baseline solver, while closely reproducing the ground truth solution ($256 \times 128$). 

The neural flux model is trained up to the final time $t = 0.2$.
Figure~\ref{fig:shock_cylinder_inlet_nf_line} compares the temporal evolution of the predicted density, velocity and pressure profiles along the $y$-direction with the reference solution and baseline results. 
At $t=0.18$ and $t=0.2$ along the line $x=0.29$ (indicated by the red dashed line in figure~\ref{fig:shock_cylinder_inlet_nf}(b)), the predictions of the neural flux model are in closer agreement with the ground truth, whereas the baseline results exhibit noticeable discrepancies. 

\section{Conclusions}
\label{sec:conclusion}
%
%
This work introduces JAX-Shock, a fully differentiable, GPU-accelerated, high-order shock-capturing neural solver for compressible flows involving complex shock-solid interactions.
By integrating WENO reconstruction, an HLLC approximate Riemann solver, and an immersed boundary method within JAX's automatic-differentiation framework, the solver enables end-to-end gradient propagation through nonlinear shock dynamics and solid-fluid interfaces--capabilities that remain largely inaccessible to traditional CFD tools.

The solver achieves high fidelity for both the Sod problem and a moving shock impinging on a cylinder. 
Benchmarking on multiple hardware architectures demonstrates excellent speedup and scalability on modern GPUs, with the NVIDIA H100 delivering the best overall performance. 

A key contribution of the framework is the neural flux module, which augments the numerical flux with data-driven corrections while preserving the stability and physical structure of the underlying shock-capturing scheme.
This hybrid formulation substantially improves coarse-grid prediction accuracy, enabling high-resolution feature recovery without incurring the full cost of fine-mesh simulation.
The solver supports multi-step differentiable unrolling, sequence-to-sequence learning, and reverse-mode inference, thereby providing a unified avenue for parameter estimation and design optimization in regimes dominated by strong discontinuities. 

The results demonstrate that JAX-Shock achieves high fidelity in canonical shock problems and challenging shock-solid configurations, while offering significant flexibility absent in purely data-driven or traditional physics-only approaches.
The framework establishes a foundation for next-generation differentiable fluid solvers capable of coupling physics-based simulation with modern machine learning methodologies. 

\section*{Acknowledgements}
This research was supported by the Division of Research and Innovation Partnership Commitments (RIPS) at Northern Illinois University. 


\section*{Data availability}
The data that support the findings of the present work are available from the corresponding author upon reasonable request. 

\section*{References}
\bibliographystyle{unsrtnat}
\bibliography{refs}

@article{Zhang_2022a,
  title={Direct numerical simulation of compressible interfacial multiphase flows using a mass--momentum--energy consistent volume-of-fluid method},
  author={Zhang, Bo and Boyd, Bradley and Ling, Yue},
  journal={Comput.~Fluids},
  volume={236},
  pages={105267},
  year={2022},
  publisher={Elsevier}
}

@article{Sod_1978a,
  title={A survey of several finite difference methods for systems of nonlinear hyperbolic conservation laws},
  author={Sod, Gary A},
  journal={J.~Comput.~Phys.},
  volume={27},
  number={1},
  pages={1--31},
  year={1978},
  publisher={Elsevier}
}

@article{Li_2021a,
  title={Kohn-Sham equations as regularizer: Building prior knowledge into machine-learned physics},
  author={Li, Li and Hoyer, Stephan and Pederson, Ryan and Sun, Ruoxi and Cubuk, Ekin D and Riley, Patrick and Burke, Kieron},
  journal={Phys.~Rev.~Lett.},
  volume={126},
  number={3},
  pages={036401},
  year={2021},
  publisher={APS}
}

@article{Xue_2023a,
  title={JAX-FEM: A differentiable GPU-accelerated 3D finite element solver for automatic inverse design and mechanistic data science},
  author={Xue, Tianju and Liao, Shuheng and Gan, Zhengtao and Park, Chanwook and Xie, Xiaoyu and Liu, Wing Kam and Cao, Jian},
  journal={Comput.~Phys.~Commun.},
  volume={291},
  pages={108802},
  year={2023},
  publisher={Elsevier}
}

@article{Schoenholz_2021a,
  title={JAX, MD A framework for differentiable physics},
  author={Schoenholz, Samuel S and Cubuk, Ekin D},
  journal={J. Stat. Mech.: Theory Exp.},
  volume={2021},
  number={12},
  pages={124016},
  year={2021},
  publisher={IOP Publishing}
}

@article{Bezgin_2023a,
  title={JAX-Fluids: A fully-differentiable high-order computational fluid dynamics solver for compressible two-phase flows},
  author={Bezgin, Deniz A and Buhendwa, Aaron B and Adams, Nikolaus A},
  journal={Comput.~Phys.~Commun.},
  volume={282},
  pages={108527},
  year={2023},
  publisher={Elsevier}
}

@article{List_2022a,
  title={Learned turbulence modelling with differentiable fluid solvers: physics-based loss functions and optimisation horizons},
  author={List, Bj{\"o}rn and Chen, Li-Wei and Thuerey, Nils},
  journal={J.~Fluid Mech.},
  volume={949},
  pages={A25},
  year={2022},
  publisher={Cambridge University Press}
}

@article{Sirignano_2020a,
  title={DPM: A deep learning PDE augmentation method with application to large-eddy simulation},
  author={Sirignano, Justin and MacArt, Jonathan F and Freund, Jonathan B},
  journal={J.~Comput.~Phys.},
  volume={423},
  pages={109811},
  year={2020},
  publisher={Elsevier}
}

@inproceedings{Zhang_2023c,
  Address = {Washington, DC, USA},
  title={A Differentiable Hybrid Neural Solver for Efficient Simulation of Cavitating Flows},
  author={Zhang, Bo and Fan, Xiantao and Wang, Jian-Xun},
  Booktitle={Bulletin of the American Physical Society},
  year={2023},
  publisher={APS}
}

@inproceedings{Zhang_2024a,
	Address = {Anaheim, CA, USA},
	Author = {Zhang, Bo and Shang, Wenjie and Panda, Jyoti and Zhou, Jiahang and Wang, Jianxun and Luo, Tengfei},
	Booktitle = {ASME 2024 Heat Transfer Summer Conference},
	Title = {JAX-BTE: A Differentiable Hybrid Neural Solver for Deep Learning Accelerated Thermal Modeling of Nanoelectronics},
	Year = {2024}}

@article{Kochkov_2021a,
  title={Machine learning--accelerated computational fluid dynamics},
  author={Kochkov, Dmitrii and Smith, Jamie A and Alieva, Ayya and Wang, Qing and Brenner, Michael P and Hoyer, Stephan},
  journal={Proc.~Natl.~Acad.~Sci.~U.S.A.},
  volume={118},
  number={21},
  pages={e2101784118},
  year={2021},
  publisher={National Academy of Sciences}
}

@article{Zhang_2025a,
  title={Banach neural operator for Navier-Stokes equations},
  author={Zhang, Bo},
  journal={Phys.~Fluids},
  volume={37},
  number={8},
  pages = {086166},
  year={2025},
  publisher={AIP Publishing}
}

@misc{jax2018github,
  author       = {Bradbury, James and Frostig, Roy and Hawkins, Peter and Johnson, Matthew J. and Leary, Chris and Maclaurin, Dougal and Necula, George and Paszke, Adam and VanderPlas, Jake and Wanderman-Milne, Skye and Zhang, Qiao},
  title        = {{JAX: composable transformations of Python+NumPy programs}},
  howpublished = {\url{http://github.com/google/jax}},
  year         = {2018}
}

@article{Paszke_2019a,
  title={Pytorch: An imperative style, high-performance deep learning library},
  author={Paszke, Adam and Gross, Sam and Massa, Francisco and Lerer, Adam and Bradbury, James and Chanan, Gregory and Killeen, Trevor and Lin, Zeming and Gimelshein, Natalia and Antiga, Luca and others},
  journal={Advances in neural information processing systems},
  volume={32},
  year={2019}
}

@inproceedings{Abadi_2016a,
  title={TensorFlow: a system for Large-Scale machine learning},
  author={Abadi, Mart{\'\i}n and Barham, Paul and Chen, Jianmin and Chen, Zhifeng and Davis, Andy and Dean, Jeffrey and Devin, Matthieu and Ghemawat, Sanjay and Irving, Geoffrey and Isard, Michael and others},
  booktitle={12th USENIX symposium on operating systems design and implementation (OSDI 16)},
  pages={265--283},
  year={2016}
}

@unpublished{Innes_2019a,
  title={A differentiable programming system to bridge machine learning and scientific computing},
  author={Innes, Mike and Edelman, Alan and Fischer, Keno and Rackauckas, Chris and Saba, Elliot and Shah, Viral B and Tebbutt, Will},
  note={arXiv preprint arXiv:1907.07587},
  year={2019}
}

@article{Chen_2025a,
  title={High-order hybrid essentially non-oscillatory spectral difference method for hyperbolic conservation laws on unstructured curved quadrilateral grids},
  author={Chen, Kuangxu and Liang, Chunlei},
  journal={Phys.~Fluids},
  volume={37},
  number={7},
  pages = {076107},
  year={2025},
  publisher={AIP Publishing}
}

@article{Olson_2003a,
  title={Preheat Effects on Shock Propagation in Indirect-Drive Inertial Confinement Fusion Ablator Materials},
  author={Olson, Richard Edward and Leeper, RJ and Nobile, A and Oertel, JA},
  journal={Phys. Rev. Lett.},
  volume={91},
  number={23},
  pages={235002},
  year={2003},
  publisher={APS}
}

@article{Muller_2020a,
  title={Hydrodynamics of core-collapse supernovae and their progenitors},
  author={M{\"u}ller, Bernhard},
  journal={Living Rev. Comput. Astrophys.},
  volume={6},
  number={1},
  pages={3},
  year={2020},
  publisher={Springer}
}

@article{Xiao_2018a,
  title={Hypersonic shock wave interactions on a V-shaped blunt leading edge},
  author={Xiao, Fengshou and Li, Zhufei and Zhang, Zhiyu and Zhu, Yujian and Yang, Jiming},
  journal={AIAA J.},
  volume={56},
  number={1},
  pages={356--367},
  year={2018},
  publisher={American Institute of Aeronautics and Astronautics}
}

@article{Kumar_2025a,
  title={Instability of isolator shocks to fuel flow rate modulations in a strut-stabilised scramjet combustor},
  author={Kumar, R and Ghosh, A},
  journal={Aeronaut. J.},
  volume={129},
  number={1331},
  pages={42--62},
  year={2025},
  publisher={Cambridge University Press}
}

@article{Tembhurnikar_2025a,
  title={CFD Analysis of Weapon Bay Missile Ejection at Various Mach Numbers},
  author={Tembhurnikar, P and Sarwar, MD G and Sahoo, D},
  journal={Fluid Dyn.},
  volume={60},
  number={1},
  pages={14},
  year={2025},
  publisher={Springer}
}

@article{Zhang_2023a,
  title={Nonlinear mode decomposition via physics-assimilated convolutional autoencoder for unsteady flows over an airfoil},
  author={Zhang, Bo},
  journal={Phys.~Fluids},
  volume={35},
  number={9},
  pages = {095115},
  year={2023},
  publisher={AIP Publishing}
}

@article{Zhang_2023b,
  title={Airfoil-based convolutional autoencoder and long short-term memory neural network for predicting coherent structures evolution around an airfoil},
  author={Zhang, Bo},
  journal={Comput.~Fluids},
  volume={258},
  pages={105883},
  year={2023}
}

@article{Raissi_2019a,
  title={Physics-informed neural networks: A deep learning framework for solving forward and inverse problems involving nonlinear partial differential equations},
  author={Raissi, Maziar and Perdikaris, Paris and Karniadakis, George E},
  journal={J.~Comput.~Phys.},
  volume={378},
  pages={686--707},
  year={2019},
  publisher={Elsevier}
}

@article{Popinet_2009a,
	Author = {Popinet, S.},
	Journal = {J.~Comput.~Phys.},
	Number = {16},
	Pages = {5838--5866},
	Title = {An accurate adaptive solver for surface-tension-driven interfacial flows},
	Volume = {228},
	Year = {2009}}

@article{Mittal_2005a,
	Author = {Mittal, R. and Iaccarino, G.},
	Journal = {Annu.~Rev.~Fluid Mech.},
	Pages = {239-261},
	Title = {Immersed boundary methods},
	Volume = {37},
	Year = {2005}}

@article{Johnsen_2006a,
	Author = {Johnsen, E. and Colonius, T.},
	Journal = {J.~Comput.~Phys.},
	Pages = {715-732},
	Title = {{Implementation of WENO schemes in compressible multicomponent flow problems}},
	Volume = {219},
	Year = {2006}}

\end{document}